# (2 +1)-dimensional Duffin-Kemmer-Petiau oscillator under a magnetic field in the presence of a minimal length in the noncommutative space

Bing-Qian Wang, Zheng-Wen Long, Chao-Yun Long and Shu-Rui Wu

College of Physics, Guizhou University, Guiyang 550025, China

Correspondence should be addressed to Zheng-Wen Long; zwlong@gzu.edu.cn

Using the momentum space representation, we study the (2 +1)-dimensional Duffin-Kemmer-Petiau oscillator for spin 0 particle under a magnetic field in the presence of a minimal length in the noncommutative space. The explicit form of energy eigenvalues are found, the wave functions and the corresponding probability density are reported in terms of the Jacobi polynomials. Additionally, we also discuss the special cases and depict the corresponding numerical results.

## 1. Introduction

The interest in the study of the minimal length uncertainty relation combination with the noncommutative space commutation relations in nonrelativistic wave equation and relativistic wave equation has drawn much attention [1-2] in recent years. Motivated by string theory, loop quantum gravity and quantum geometry [3-15], the modification of the ordinary uncertainty relation has become an appealing case of research. In the so-called minimal length formulation, Kempf et al [16-18] have shown that the minimal length can be introduced as an additional uncertainty in position measurement, so that the usual canonical commutation relation between position and momentum operator is substituted by $[x, p] = i\hbar(1 + \beta p^2)$, where $\beta$ is a small positive parameter called the deformation parameter. Recently, various topics have been studied in connection with the minimal length uncertainty relation [19-28]. On the other hand, the issue of noncommutative (NC) quantum mechanics has also been extensively discussed. The noncommutativity of space-time coordinates was first introduced by Snyder [29] aiming to improve the problem of infinite self-energy in quantum field theory, and the noncommutative geometry has been put forward because of the discovery in string theory and matrix model of M-theory [30]. Recently, various aspects of both NC classical [31] and quantum[32] mechanics have been extensively studied devoted to exploring the role of NC parameter in the physical observables [33-40]. Recently, the effect of the quantum gravity on the quantum mechanics by modifying the basic commutators among the canonical variables has been an attractive topic, therefore the combination of the minimal length uncertainty relation and the noncommutative space commutation relations is a colorful problem. In the present work, we are interesting to study the two sectors in the framework of the relativistic DKP equation, and analyse the effects of them on the energy spectrum and the corresponding wave functions.

The organization of this work is as follows. In Section 2, we consider the DKP oscillator for spin 0 particle in the presence of a minimal length in the NC space. In Section 3, we study the problem under a magnetic field. In Section 4, we discuss some special cases of the solutions to check the validity of our results. Finally, the work is summarized in last section.

**2. The (2 +1)-DKP Oscillator for Spin 0 Particle in the presence of the minimal length in NC space**

In NC space, the canonical variables satisfy the following commutation relations:

$$\left[x_i^{(NC)}, x_j^{(NC)}\right] = i\theta_{ij}, \quad \left[P_i^{(NC)}, P_j^{(NC)}\right] = 0, \quad \left[x_i^{(NC)}, P_j^{(NC)}\right] = i\hbar\delta_{ij}, (i,j = 1,2), \quad (1)$$

with $\theta_{ij} = \epsilon_{ij}\theta$ is the antisymmetric NC parameter, representing the noncommutativity of the space, and $x_i^{(NC)}$, $P_i^{(NC)}$ are the coordinate and momentum operators in the NC space. By replacing the normal product with star product, the DKP equation in commutation space will change into the DKP equation in NC space as

$$H(P,x) * \psi(x) = E\psi(x). \quad (2)$$

Usually, one way to deal with the problem of NC space is via the star product or Moyal Weyl product on the commutative space functions:

$$(f * g)(x) = \exp\left[\frac{i}{2}\theta_{ij}\partial_i^x \partial_j^y\right] f(x)g(y)|_{y\to x}, \quad (3)$$

where $f(x)$ and $g(y)$ are arbitrary infinitely differentiable functions.

Then the Moyal-Weyl product can be replaced by a Bopp shift of the form

$$x_i^{(NC)} = \hat{x}_i - \frac{1}{2\hbar}\theta_{ij}\widehat{P}_j, \quad P_i^{(NC)} = \widehat{P}_i, \quad (4)$$

therefore in the two dimensional NC space, Eq. (4) can be expressed as

$$x^{(NC)} = \hat{x} - \frac{\theta}{2\hbar}\widehat{P}_y, \quad y^{(NC)} = \hat{y} + \frac{\theta}{2\hbar}\widehat{P}_x, \quad P_x^{(NC)} = \widehat{P}_x, \quad P_y^{(NC)} = \widehat{P}_y, \quad (5)$$

where $\hat{x}$, $\hat{y}$, $\widehat{P}_x$ and $\widehat{P}_y$ are the position and momentum operators in the usual quantum mechanics respectively, which satisfy the canonical Heisenberg commutation relations.

Thus, the relativistic DKP equation for a free boson of mass $m$ is given by [41-42]

$$[c\vec{\beta} \cdot \vec{P} + mc^2]\psi' = i\hbar\beta^0 \frac{\partial}{\partial t}\psi', \tag{6}$$

where $\vec{\beta} = (\beta^1, \beta^2, \beta^3)$ and $\beta^0$ is the DKP matrices which meet the following algebra relation:

$$\beta^\mu \beta^\nu \beta^\lambda + \beta^\lambda \beta^\nu \beta^\mu = g^{\mu\nu}\beta^\lambda + g^{\lambda\nu}\beta^\mu, \tag{7}$$

with $g^{\mu\nu} = \text{diag}(1, -1, -1, -1)$ is the metric tensor in Minkowski space. For spin 0 particle, $\beta^i$ are $5 \times 5$ matrices expressed as

$$\beta^0 = \begin{bmatrix} I_0 & \bar{0} \\ \bar{0}^T & \bar{\sigma} \end{bmatrix}, \quad \beta^i = \begin{bmatrix} \hat{0} & k_i \\ -(k_i)^T & \bar{\sigma} \end{bmatrix}, i = 1,2,3, \tag{8}$$

with $\bar{\sigma}$, $\bar{0}$ and $\hat{0}$ being $3 \times 3$, $2 \times 3$, and $2 \times 2$ zero matrices, respectively. Other matrices in equation (8) are given as follows:

$$I_0 = \begin{bmatrix} 0 & 1 \\ 1 & 0 \end{bmatrix}, k_1 = \begin{bmatrix} -1 & 0 & 0 \\ 0 & 0 & 0 \end{bmatrix}, k_2 = \begin{bmatrix} 0 & -1 & 0 \\ 0 & 0 & 0 \end{bmatrix}, k_3 = \begin{bmatrix} 0 & 0 & -1 \\ 0 & 0 & 0 \end{bmatrix}. \tag{9}$$

The DKP oscillator is introduced by using substitution of the momentum operator $\vec{P}$ with $(\vec{P} \to \vec{P} - im\omega\eta^0\vec{r})$, where the additional term is linear in r, w is the oscillator frequency, and $\eta^0 = 2(\beta^0)^2 - I$ with $(\eta^0)^2 = I$.

It is easy to get the 2D DKP oscillator from the above equation:

$$[c\vec{\beta} \cdot (\vec{P} - im\omega\eta^0\vec{r}) + mc^2]\psi = i\hbar\beta^0 \frac{\partial}{\partial t}\psi. \tag{10}$$

Given this situation above, considering the NC formalism and via the Bopp shift (4-5), the (2+1)-dimensional DKP oscillator equation in NC space becomes

$$\left[c\beta^1\left(P_x^{(NC)} - imw\eta^0 x^{(NC)}\right) + c\beta^2\left(P_y^{(NC)} - imwy^{(NC)}\right) + mc^2\right]\psi = \beta^0 E. \tag{11}$$

For a boson of spin 0, the spinor $\psi$ is a vector with five components [43, 44], which reads

$$\psi = (\psi_1, \psi_2, \psi_3, \psi_4, \psi_5)^T. \tag{12}$$

Substituting $\psi$ into (11) we have

$$-mc^2\psi_1 + E\psi_2 + c\left(P_x^{(NC)} + imwx^{(NC)}\right)\psi_3 + c\left(P_y^{(NC)} + imwy^{(NC)}\right)\psi_4 = 0,$$
$$E\psi_1 - mc^2\psi_2 = 0,$$

$$c\left(P_x^{(NC)} - imwx^{(NC)}\right)\psi_1 + mc^2\psi_3 = 0,$$

$$c\left(P_y^{(NC)} - imwy^{(NC)}\right)\psi_1 + mc^2\psi_4 = 0,$$

$$mc^2\psi_5 = 0. \tag{13}$$

Combination of (13) gives

$$\left\{\left[\widehat{P}_x + imw\left(\hat{x} - \frac{\theta}{2\hbar}\widehat{P}_y\right)\right]\left[\widehat{P}_x - imw\left(\hat{x} - \frac{\theta}{2\hbar}\widehat{P}_y\right)\right] + \left[\widehat{P}_y + imw\left(\hat{y} + \frac{\theta}{2\hbar}\widehat{P}_x\right)\right]\left[\widehat{P}_y - imw\left(\hat{y} + \frac{\theta}{2\hbar}\widehat{P}_x\right)\right] + m^2c^2 - \frac{E^2}{c^2}\right\}\psi_1 = 0. \tag{14}$$

In addition, in the minimal length formalism, the Heisenberg algebra is given by

$$\left[\hat{x}_i, \widehat{P}_j\right] = i\hbar\delta_{ij}(1 + \alpha P^2), \tag{15}$$

where α is the minimal length positive parameter. Moreover, in the momentum space, the position vector and momentum vector can be expressed as

$$\hat{x}_i = i\hbar\left[(1 + \alpha P^2)\frac{\partial}{\partial p_i}\right], \widehat{P}_i = P_i, \tag{16}$$

substituting (15) and (16) into (14) we have

$$\left[\left(1 + \frac{m^2w^2\theta^2}{4\hbar^2}\right)(P_x^2 + P_y^2) + m^2w^2(x^2 + y^2) - 2mw\hbar(1 + \alpha P^2) - \frac{m^2w^2\theta}{\hbar}(1 + \alpha P^2)L_z + m^2c^2 - \frac{E^2}{c^2}\right]\psi_1 = 0. \tag{17}$$

Now, in order to solve Eq. (17), an auxiliary wave function is defined as $\psi_1(P, \vartheta) = e^{i\ell\vartheta}\phi(P)$, and for the sake of simplicity, we bring the problem into the polar coordinates, recalling that

$$P_x = P\cos\vartheta, P_y = P\sin\vartheta,$$

$$\frac{\partial}{\partial P_x} = \cos\vartheta\frac{\partial}{\partial P} - \frac{\sin\vartheta}{P}\frac{\partial}{\partial\vartheta}, \frac{\partial}{\partial P_y} = \sin\vartheta\frac{\partial}{\partial P} + \frac{\cos\vartheta}{P}\frac{\partial}{\partial\vartheta}, \tag{18}$$

then Eq. (17) becomes

$$\left\{(1 + \alpha P^2)^2\frac{\partial^2}{\partial P^2} + \left[\frac{1}{P}(1 + \alpha P^2)^2 + 2\alpha P(1 + \alpha P^2)\right]\frac{\partial}{\partial P} - \frac{\ell^2}{P^2}(1 + \alpha P^2)^2 + \frac{-\left(1 + \frac{m^2w^2\theta^2}{4\hbar^2}\right) + 2mw\alpha\hbar + m^2w^2\theta\alpha\ell}{m^2w^2\hbar^2}P^2 + \left(\frac{E^2 - m^2c^4}{c^2} + 2mw\hbar + m^2w^2\theta\ell\right)/m^2w^2\hbar^2\right\}\phi(P) = 0,$$

(19)

with the help of a variable transformation

$$q = \frac{1}{\sqrt{\alpha}}\tan^{-1}(\sqrt{\alpha}P), \tag{20}$$

which will map the variable $P \in (0, \infty)$ to $q \in \left(0, \frac{\pi}{2\sqrt{\alpha}}\right)$, we simplify Eq. (19) to

$$\left[\frac{\partial^2}{\partial q^2} + \left(\frac{\upsilon}{\mu} + \frac{\mu}{\upsilon}\right)\sqrt{\alpha}\frac{\partial}{\partial q} - \alpha\ell^2\frac{\upsilon^2}{\mu^2} + \alpha k'\frac{\mu^2}{\upsilon^2} + \alpha\varepsilon'\right]\phi(q) = 0, \tag{21}$$

where

$$\mu = \sin(\sqrt{\alpha}q), \qquad \upsilon = \cos(\sqrt{\alpha}q),$$

$$k' = -\ell^2 + \frac{-\left(1+\frac{m^2w^2\theta^2}{4\hbar^2}\right)+2mw\alpha\hbar+m^2w^2\theta\alpha\ell}{\alpha^2 m^2 w^2 \hbar^2}, \quad \varepsilon' = \frac{\frac{E^2-m^2c^4}{c^2}+2mw\hbar+m^2w^2\theta\ell}{\alpha m^2 \omega^2 \hbar^2} - 2\ell^2. \tag{22}$$

Next, for convenience, another auxiliary function is introduced by $\phi(q) = \upsilon^\lambda F(q)$, with $\lambda$ being a constant to be determined. Thus we have

$$\left\{\frac{\partial^2}{\partial q^2} + \left(\frac{\upsilon}{\mu} + (1-2\lambda)\frac{\mu}{\upsilon}\right)\sqrt{\alpha}\frac{\partial}{\partial q} - \alpha\ell^2\frac{\upsilon^2}{\mu^2} + \alpha[k' + \lambda(\lambda-1) - \lambda]\frac{\mu^2}{\upsilon^2} + \alpha(\varepsilon' - 2\lambda)\right\}F(q) = 0.$$

(23)

Here, in order to simplify above mathematical expression, we select $k' + \lambda(\lambda - 1) - \lambda = 0$, then it leads to the following expression of $\lambda$

$$\lambda = 1 + \sqrt{1-k'}, \quad \lambda' = 1 - \sqrt{1-k'}. \tag{24}$$

Since the second solution leads to a non physically acceptable wave function, then equation (23) turns into

$$\left[\frac{\partial^2}{\partial q^2} + \left(\frac{\upsilon}{\mu} + (1-2\lambda)\frac{\mu}{\upsilon}\right)\sqrt{\alpha}\frac{\partial}{\partial q} - \alpha\ell^2\frac{\upsilon^2}{\mu^2} + \alpha(\varepsilon' - 2\lambda)\right]F(q) = 0. \tag{25}$$

Then with the help of another auxiliary function $F(q) = \mu^\ell \zeta(q)$, thus eq. (25) reads

$$\left\{\frac{\partial^2}{\partial q^2} + \left[(2\ell+1)\frac{\upsilon}{\mu} + (1-2\lambda)\frac{\mu}{\upsilon}\right]\sqrt{\alpha}\frac{\partial}{\partial q} + \alpha(\varepsilon' - 2\lambda - 2\lambda\ell)\right\}\zeta(q) = 0. \tag{26}$$

Now we make a variable transformation by demanding $z = 2\mu^2 - 1$, where the variable interval is $z \in (-1, 1)$, then one can obtain

$$\left\{(1-z^2)\frac{\partial^2}{\partial z^2}+[(\ell+1-\lambda)-(\ell+1+\lambda)z]\frac{\partial}{\partial z}+\frac{1}{4}(\varepsilon'-2\lambda-2\lambda\ell)\right\}\zeta(z)=0. \tag{27}$$

It is important to point out that the wave function we used here will be regular at $z=\pm 1$ on the condition that $\zeta(z)$ is a polynomial, which is obtained by imposing the following constraint $\frac{1}{4}(\varepsilon'-2\lambda-2\lambda\ell)=n(n+\ell+\lambda)$, with n being a non-negative integer. Then eq. (27) turns into

$$\left\{(1-z^2)\frac{\partial^2}{\partial z^2}+[(\ell+1-\lambda)-(\ell+1+\lambda)z]\frac{\partial}{\partial z}+n(n+\ell+\lambda)\right\}\zeta(z)=0, \tag{28}$$

whose solution can be written in terms of Jacobi polynomials as $\zeta(z)=P_n^{(a,b)}(z)$, where $a=\lambda-1, b=\ell$. In this case, the energy eigenvalue of the system can be expressed as

$$\varepsilon''=2\ell^2+4n^2+4n\ell+[4n+2(\ell+1)](1+\sqrt{1-k'}), \tag{29}$$

with $\varepsilon''=\frac{E^2+2m\omega\hbar c^2-m^2c^4}{\alpha m^2\omega^2\hbar^2 c^2}$.

Therefore the energy eigenvalues can be derived from eq. (29) as

$$E_{n\ell}=\pm mc^2\left\{(4n^2+4n\ell+2\ell^2)\frac{\alpha\omega^2\hbar^2}{c^2}-\frac{2\hbar\omega}{mc^2}-\frac{w^2\theta\ell}{c^2}+1+\frac{2\alpha\omega^2\hbar^2}{c^2}(2n+\ell+1)\left[1+\sqrt{1+\ell^2+\frac{\left(1+\frac{m^2w^2\theta^2}{4\hbar^2}\right)-2mw\alpha\hbar-m^2w^2\theta\alpha\ell}{\alpha^2m^2w^2\hbar^2}}\right]\right\}^{\frac{1}{2}}, n,\ell=0,1,2,\cdots, \tag{30}$$

furthermore, the wave function may be expressed by

$$\psi_1(P)=Ne^{i\ell\vartheta}\left(\frac{1}{1+\alpha P^2}\right)^{\frac{\lambda}{2}}\left(\frac{\alpha P^2}{1+\alpha P^2}\right)^{\frac{\ell}{2}}P_n^{(a,b)}\left(\frac{2\alpha P^2}{1+\alpha P^2}-1\right), n,\ell=0,1,2\cdots, \tag{31}$$

Then the wave function of the system is

$$\psi=N\begin{pmatrix}1\\ E_{n\ell}/mc^2\\ -\left[\left(\cos\vartheta+i\frac{\theta}{2\hbar}mw\sin\vartheta\right)P+\hbar mw(1+\alpha P^2)\left(\cos\vartheta\frac{\partial}{\partial P}-\frac{\sin\vartheta}{P}\frac{\partial}{\partial\vartheta}\right)\right]\Big/mc\\ -\left[\left(\sin\vartheta-i\frac{\theta}{2\hbar}mw\cos\vartheta\right)P+m\omega\hbar(1+\alpha P^2)\left(\sin\vartheta\frac{\partial}{\partial P}+\frac{\cos\vartheta}{P}\frac{\partial}{\partial\vartheta}\right)\right]\Big/mc\\ 0\end{pmatrix}\psi_1(P,\vartheta).$$

(32)

Now the Jacobi polynomial [45] is employed to obtain the other components wave:

$$\frac{dP_n^{(a,b)}(t)}{dt} = \frac{1}{2}(n+a+b+1)P_{n-1}^{(a+1,b+1)}(t), \tag{33}$$

we finally have

$$\psi_2(P,\vartheta) = \frac{E_{n\ell}}{mc^2}\psi_1(P,\vartheta) = N\frac{E_{n\ell}}{mc^2}e^{i\ell\vartheta}\left(\frac{1}{1+\alpha P^2}\right)^{\frac{\lambda}{2}}\left(\frac{\alpha P^2}{1+\alpha P^2}\right)^{\frac{\ell}{2}}P_n^{(a,b)}\left(\frac{\alpha P^2-1}{1+\alpha P^2}\right),$$

$$\psi_3(P,\vartheta) = -\frac{\left[\left(\cos\vartheta + i\frac{\theta}{2\hbar}mw\sin\vartheta\right)P + \hbar mw(1+\alpha P^2)\left(\cos\vartheta\frac{\partial}{\partial P} - \frac{\sin\vartheta}{P}\frac{\partial}{\partial\vartheta}\right)\right]}{mc}\psi_1(P,\vartheta)$$

$$= -Ne^{i\ell\vartheta}\left(\frac{1}{1+\alpha P^2}\right)^{\frac{\lambda}{2}}\left(\frac{\alpha P^2}{1+\alpha P^2}\right)^{\frac{\ell}{2}}\frac{1}{mc}\left\{\left[\left(\cos\vartheta + i\frac{\theta}{2\hbar}mw\sin\vartheta\right)P + \hbar mw\cos\vartheta\left(-\lambda\alpha P + \frac{\ell}{P}\right) - \right.\right.$$

$$\left.i\hbar mw\ell(1+\alpha P^2)\frac{\sin\vartheta}{P}\right]P_n^{(\lambda-1,\ell)} + \hbar mw\cos\vartheta\frac{2\alpha P}{1+\alpha P^2}(n+\ell+\lambda)P_{n-1}^{(\lambda,\ell+1)}\right\},$$

$$\psi_4(P,\vartheta) = -\frac{\left[\left(\sin\vartheta - i\frac{\theta}{2\hbar}mw\cos\vartheta\right)P + mw\hbar(1+\alpha P^2)\left(\sin\vartheta\frac{\partial}{\partial P} + \frac{\cos\vartheta}{P}\frac{\partial}{\partial\vartheta}\right)\right]}{mc}\psi_1(P,\vartheta)$$

$$= -Ne^{i\ell\vartheta}\left(\frac{1}{1+\alpha P^2}\right)^{\frac{\lambda}{2}}\left(\frac{\alpha P^2}{1+\alpha P^2}\right)^{\frac{\ell}{2}}\frac{1}{mc}\left\{\left[\left(\sin\vartheta - i\frac{\theta}{2\hbar}mw\cos\vartheta\right)P + mw\hbar\sin\vartheta\left(-\lambda\alpha P + \frac{\ell}{P}\right) + \right.\right.$$

$$\left.i\hbar mw\ell(1+\alpha P^2)\frac{\cos\vartheta}{P}\right]P_n^{(\lambda-1,\ell)} + \hbar mw\sin\vartheta\frac{2\alpha P}{1+\alpha P^2}(n+\ell+\lambda)P_{n-1}^{(\lambda,\ell+1)}\right\},$$

$$\psi_5(P,\vartheta) = 0. \tag{34}$$

After ending this part, we determine the normalization constant N by demanding

$$\int_{-\infty}^{+\infty}\frac{d^2p}{(1+\alpha P^2)}\overline{\psi}(P)\beta^0\psi(P) = 1, \tag{35}$$

besides, according to the following property of the Jacobi polynomial

$$\int_{-1}^{+1}dt(1-t)^a(1+t)^b\left[P_n^{(a,b)}(t)\right]^2 = \frac{2^{a+b+1}\Gamma(a+n+1)\Gamma(b+n+1)}{n!(a+b+1+2n)\Gamma(a+b+n+1)}, \tag{36}$$

we obtain

$$N = \alpha^{\frac{1}{2}}\left[\frac{mc^2 n!(2n+\lambda+\ell)\Gamma(n+\lambda+\ell)}{E_{n\ell}\Gamma(n+\lambda)\Gamma(n+\ell+1)}\right]^{\frac{1}{2}}, \tag{37}$$

then one can obtain the corresponding probability density of every component given by

$$P'_i = |\int_0^\pi \int_0^\infty \bar{\psi}_i(P,\vartheta)\beta^0\psi_i(P,\vartheta)dpd\vartheta|, \ i=1,2,3,4,5. \tag{38}$$

## 3. The Problem under a magnetic field

Now, in the presence of an external magnetic field, i.e. $\vec{A} = \left(-\frac{By^{(NC)}}{2} \ \frac{Bx^{(NC)}}{2} \ 0\right)$, the Eq. (11) is transformed into

$$\left[\beta^0\tilde{E} - c\beta^1\left(P_x^{(NC)} + \frac{eBy^{(NC)}}{2c} - im\omega\eta^0 x^{(NC)}\right) - c\beta^2\left(P_y^{(NC)} - \frac{eBx^{(NC)}}{2c} - im w\eta^0 y^{(NC)}\right) - mc^2\right]\tilde{\psi} = 0, \tag{39}$$

here the spinor $\tilde{\psi}$ is also a vector with five components which reads

$$\tilde{\psi} = (\tilde{\psi}_1, \tilde{\psi}_2, \tilde{\psi}_3, \tilde{\psi}_4, \tilde{\psi}_5)^T. \tag{40}$$

Substituting $\tilde{\psi}$ into (39) one can obtain

$$-mc^2\tilde{\psi}_1 + \tilde{E}\tilde{\psi}_2 + c\left(P_x^{(NC)} + \frac{eBy^{(NC)}}{2c} + imwx^{(NC)}\right)\tilde{\psi}_3 +$$
$$c\left(P_y^{(NC)} - \frac{eBx^{(NC)}}{2c} + imwy^{(NC)}\right)\tilde{\psi}_4 = 0,$$

$$\tilde{E}\tilde{\psi}_1 - mc^2\tilde{\psi}_2 = 0,$$

$$c\left(P_x^{(NC)} + \frac{eBy^{(NC)}}{2c} - imwx^{(NC)}\right)\tilde{\psi}_1 + mc^2\tilde{\psi}_3 = 0,$$

$$c\left(P_y^{(NC)} - \frac{eBx^{(NC)}}{2c} - imwy^{(NC)}\right)\tilde{\psi}_1 + mc^2\tilde{\psi}_4 = 0,$$

$$mc^2\tilde{\psi}_5 = 0. \tag{41}$$

Simplifying Eq. (41) gives

$$\left\{\left[\left(1 + \frac{eB\theta}{4\hbar c}\right)\widehat{P}_x + \frac{eB\hat{y}}{2c} + imw\left(\hat{x} - \frac{\theta}{2\hbar}\widehat{P}_y\right)\right]\left[\left(1 + \frac{eB\theta}{4\hbar c}\right)\widehat{P}_x + \frac{eB\hat{y}}{2c} - imw\left(\hat{x} - \frac{\theta}{2\hbar}\widehat{P}_y\right)\right] + \left[\left(1 + \frac{eB\theta}{4\hbar c}\right)\widehat{P}_y - \frac{eB\hat{x}}{2c} + imw\left(\hat{y} + \frac{\theta}{2\hbar}\widehat{P}_x\right)\right]\left[\left(1 + \frac{eB\theta}{4\hbar c}\right)\widehat{P}_y - \frac{eB\hat{x}}{2c} - imw\left(\hat{y} + \frac{\theta}{2\hbar}\widehat{P}_x\right)\right] + \frac{m^2c^4 - \tilde{E}^2}{c^2}\right\}\tilde{\psi}_1 = 0, \tag{42}$$

and considering Eq. (16) and Eq. (42) one can obtain

$$\left\{\left\{\left(1+\frac{eB\theta}{4\hbar c}\right)^2 + \frac{m^2w^2\theta^2}{4\hbar^2} - 2\hbar\alpha\left(mw\left(1+\frac{eB\theta}{4\hbar c}\right)+\frac{eB\theta mw}{4\hbar c}\right) - \alpha\hbar\ell\left[\frac{eB}{c}\left(1+\frac{eB\theta}{4\hbar c}\right)+\frac{m^2w^2\theta}{\hbar}-\right.\right.\right.$$
$$\left.\left. 2\hbar\alpha\frac{eBmw}{c}\right]\right\}P^2 + \left[m^2w^2 + \left(\frac{eB}{2c}\right)^2\right](x^2+y^2) + \frac{m^2c^4-\widetilde{E}^2}{c^2} - 2\hbar\left[mw\left(1+\frac{eB\theta}{4\hbar c}\right)+\frac{eB\theta mw}{4\hbar c}\right] -$$
$$\left. \hbar\ell\left[\frac{eB}{c}\left(1+\frac{eB\theta}{4\hbar c}\right)+\frac{m^2w^2\theta}{\hbar}-2\hbar\alpha\frac{eBmw}{c}\right]\right\}\tilde{\psi}_1(P,\vartheta) = 0. \qquad (43)$$

Now, by a series of analogical algebraic operations, and for the sake of simplification, we just give the results:

the energy eigenvalues of the system are

$$\widetilde{E}_{\tilde{n}\ell} = \pm mc^2\left\{\frac{\alpha\hbar^2}{m^2c^2}(4\tilde{n}^2+4\tilde{n}\ell+2\ell^2)\left[m^2w^2+\left(\frac{eB}{2c}\right)^2\right] - \frac{2\hbar}{m^2c^2}\left[mw\left(1+\frac{eB\theta}{4\hbar c}\right)+\frac{eB\theta mw}{4\hbar c}\right] - \frac{\hbar\ell}{m^2c^2}\left[\frac{eB}{c}\left(1+\frac{eB\theta}{4\hbar c}\right)+\frac{m^2w^2\theta}{\hbar}-\frac{2\hbar\alpha eBmw}{c}\right] + 1 + \frac{2\hbar^2}{m^2c^2}(2\tilde{n}+\ell+1)\left(\alpha+\sqrt{\alpha^2-\alpha^2\widetilde{K}}\right)\left(m^2w^2+\left(\frac{eB}{2c}\right)^2\right)\right\}^{\frac{1}{2}}, \quad \tilde{n}, \ell = 0,1,2,\cdots \qquad (44)$$

where

$$\widetilde{K} = -\ell^2 - \frac{\left\{\left(1+\frac{eB\theta}{4\hbar c}\right)^2+\frac{m^2w^2\theta^2}{4\hbar^2}-2\hbar\alpha\left(mw\left(1+\frac{eB\theta}{4\hbar c}\right)+\frac{eB\theta mw}{4\hbar c}\right)-\alpha\hbar\ell\left[\frac{eB}{c}\left(1+\frac{eB\theta}{4\hbar c}\right)+\frac{m^2w^2\theta}{\hbar}-\frac{2\hbar\alpha eBmw}{c}\right]\right\}}{\alpha^2\hbar^2\left[m^2w^2+\left(\frac{eB}{2c}\right)^2\right]}. \qquad (45)$$

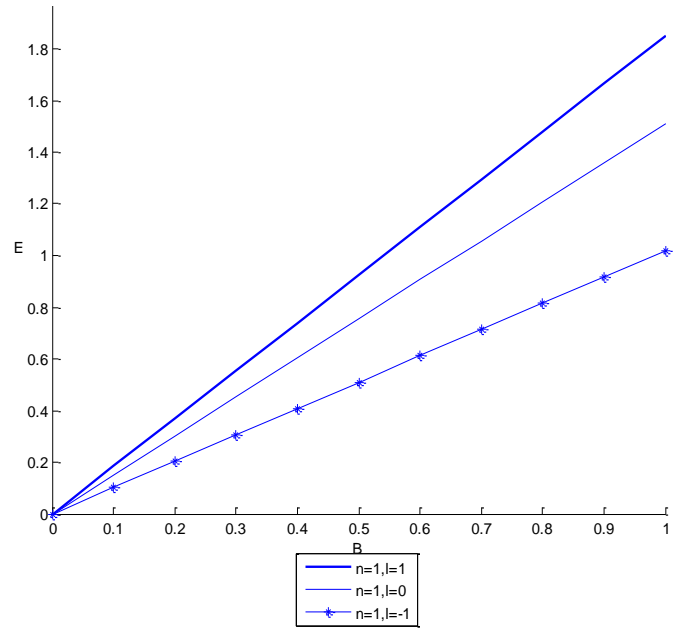

Figure 1: The energy eigenvalues versus $B$ ($\theta = 0.00003$, m = 1).

The wave function can be expressed as

$$\tilde{\psi}_1(P,\vartheta) = \tilde{N}e^{i\ell\vartheta}\left(\frac{1}{1+\alpha P^2}\right)^{\frac{\lambda'}{2}}\left(\frac{\alpha P^2}{1+\alpha P^2}\right)^{\frac{\ell}{2}}P_{\tilde{n}}^{(a',b')}\left(\frac{2\alpha P^2}{1+\alpha P^2}-1\right), \tilde{n}, \ell = 0,1,2,3,4\cdots, \quad (46)$$

thus the wave function of the system is

$$\tilde{\psi} = \tilde{N}\begin{pmatrix} 1 \\ E_{\tilde{n}\ell}/mc^2 \\ -\left\{\left[\left(1+\frac{eB\theta}{4\hbar c}\right)\cos\vartheta + \frac{im w\theta}{2\hbar}\sin\vartheta\right]P + (1+\alpha P^2)\left(\frac{i\hbar eB}{2c}\frac{\partial}{\partial P_y} + \hbar mw\frac{\partial}{\partial P_x}\right)\right\}/mc \\ \left\{\left[\left(1+\frac{eB\theta}{4\hbar c}\right)\sin\vartheta - \frac{im w\theta}{2\hbar}\cos\vartheta\right]P + (1+\alpha P^2)\left(-\frac{i\hbar eB}{2c}\frac{\partial}{\partial P_x} + \hbar mw\frac{\partial}{\partial P_y}\right)\right\}/mc \\ 0 \end{pmatrix}\tilde{\psi}_1(P,\vartheta),$$

(47)

then the other components wave function of the system are

$$\tilde{\psi}_2(P,\vartheta) = \frac{E_{\tilde{n}\ell}}{mc^2}\tilde{\psi}_1(P,\vartheta) = \tilde{N}\frac{E_{\tilde{n}\ell}}{mc^2}e^{i\ell\vartheta}\left(\frac{1}{1+\alpha P^2}\right)^{\frac{\lambda'}{2}}\left(\frac{\alpha P^2}{1+\alpha P^2}\right)^{\frac{\ell}{2}}P_{\tilde{n}}^{(a',b')}\left(\frac{2\alpha P^2}{1+\alpha P^2}-1\right),$$

$$\widetilde{\psi}_3(P,\vartheta)$$
$$= -\frac{\left\{\left[\left(1+\frac{eB\theta}{4\hbar c}\right)\cos\vartheta + \frac{im w\theta}{2\hbar}\sin\vartheta\right]P + (1+\alpha P^2)\left(\frac{i\hbar eB}{2c}\frac{\partial}{\partial P_y} + \hbar mw\frac{\partial}{\partial P_x}\right)\right\}}{mc}\tilde{\psi}_1(P,\vartheta)$$

$$= -\frac{\tilde{N}e^{i\ell\vartheta}}{mc}\left(\frac{1}{1+\alpha P^2}\right)^{\frac{\lambda'}{2}}\left(\frac{\alpha P^2}{1+\alpha P^2}\right)^{\frac{\ell}{2}}\left\{\left[\left(1+\frac{eB\theta}{4\hbar c}\right)\cos\vartheta + \frac{im w\theta}{2\hbar}\sin\vartheta\right]P + \left(\frac{i\hbar eB}{2c}\sin\vartheta + \right.\right.$$

$$\left.\hbar mw\cos\vartheta\right)\left(-\lambda'\alpha P + \frac{\ell}{P}\right) - \frac{\hbar eB\ell}{2c}(1+\alpha P^2)\frac{\cos\vartheta}{P} - i\ell\hbar mw(1+\alpha P^2)\frac{\sin\vartheta}{P}\right\}P_{\tilde{n}}^{(\lambda'-1,\ell)} +$$

$$\left(\frac{i\hbar eB}{2c}\sin\vartheta + \hbar mw\cos\vartheta\right)\frac{2\alpha P}{1+\alpha P^2}(n+\ell+\lambda')P_{\tilde{n}-1}^{(\lambda',\ell-1)}\bigg\},$$

$$\tilde{\psi}_4(P,\vartheta)$$
$$= -\frac{\left\{\left[\left(1+\frac{eB\theta}{4\hbar c}\right)\sin\vartheta - \frac{im w\theta}{2\hbar}\cos\vartheta\right]P + (1+\alpha P^2)\left(-\frac{i\hbar eB}{2c}\frac{\partial}{\partial P_x} + \hbar mw\frac{\partial}{\partial P_y}\right)\right\}}{mc}\tilde{\psi}_1(P,\vartheta)$$

$$= -\frac{\tilde{N}e^{i\ell\vartheta}}{mc}\left(\frac{1}{1+\alpha P^2}\right)^{\frac{\lambda'}{2}}\left(\frac{\alpha P^2}{1+\alpha P^2}\right)^{\frac{\ell}{2}}\left\{\left[\left(1+\frac{eB\theta}{4\hbar c}\right)\sin\vartheta - \frac{im w\theta}{2\hbar}\cos\vartheta\right]P + \left(-\frac{i\hbar eB}{2c}\cos\vartheta + \right.\right.$$

$$\left.\hbar mw\sin\vartheta\right)\left(-\lambda\alpha P + \frac{\ell}{P}\right) + \left(\frac{i\hbar eB}{2c}\frac{\sin\vartheta}{P} + \hbar mw\frac{\cos\vartheta}{P}\right)i\ell(1+\alpha P^2)\right\}P_{\tilde{n}}^{(\lambda'-1,\ell)} + \left(-\frac{i\hbar eB}{2c}\cos\vartheta + \right.$$

$$\left.\hbar mw\sin\vartheta\right)\frac{2\alpha P}{1+\alpha P^2}(n+\ell+\lambda')P_{\tilde{n}-1}^{(\lambda',\ell-1)}\bigg\},$$

$$\tilde{\psi}_5(P,\vartheta) = 0. \tag{48}$$

The normalization constant N is

$$\widetilde{N} = \alpha^{\frac{1}{2}} \left[ \frac{Mc^2 \tilde{n}! (2\tilde{n}+\lambda'+\ell)\Gamma(\tilde{n}+\lambda'+\ell)}{E_{\tilde{n}\ell}\Gamma(\tilde{n}+\lambda')\Gamma(\tilde{n}+\ell+1)} \right]^{\frac{1}{2}}, \tag{49}$$

finally, the corresponding probability density of every component can be expressed as

$$P''_i = \left| \int_0^\pi \int_0^\infty \bar{\tilde{\psi}}_i(P,\vartheta) \beta^0 \tilde{\psi}_i(P,\vartheta) dp d\vartheta \right|, \text{ i=1,2,3,4,5.} \tag{50}$$

## 4. Special cases and discussions

In this section, the natural unit ($\hbar = c = m = w = 1$) is employed. Now, let us check the special cases. First, when the minimal length parameter $\alpha \to 0$, the energy relation (44) reduces to

$$\widetilde{E}_{\tilde{n}\ell} = \pm mc^2 \left\{ 1 - \frac{2\hbar}{m^2 c^2} \left[ mw\left(1 + \frac{eB\theta}{4\hbar c}\right) + \frac{eB\theta mw}{4\hbar c} \right] - \frac{\hbar\ell}{m^2 c^2} \left[ \frac{eB}{c}\left(1 + \frac{eB\theta}{4\hbar c}\right) + \frac{m^2 w^2 \theta}{\hbar} \right] + \widetilde{K}' \right\}^{\frac{1}{2}},$$
$$\tilde{n}, \ell = 0,1,2,\cdots \tag{51}$$

where

$$\widetilde{K}' = \frac{2\hbar^2}{m^2 c^2}(2\tilde{n}+\ell+1)\left(m^2 w^2 + \left(\frac{eB}{2c}\right)^2\right)\left(l^2 + \frac{\left(1+\frac{eB\theta}{4\hbar c}\right)^2 + \frac{m^2 w^2 \theta^2}{4\hbar^2}}{\hbar^2\left(m^2 w^2 + \left(\frac{eB}{2c}\right)^2\right)}\right)^{\frac{1}{2}}. \tag{52}$$

The energy has plotted the energy eigenvalues versus $B$ in Figure 1. We see that the energy E increases monotonically with the magnetic field parameter $B$ and the tendency of the spectrum can be observed for large numbers. It also shows that for one principal quantum number, the energy E increases with the increase of the azimuthal quantum number. The energy relation in the special case of $\theta = 0$, i.e. for vanishing NC parameter gives

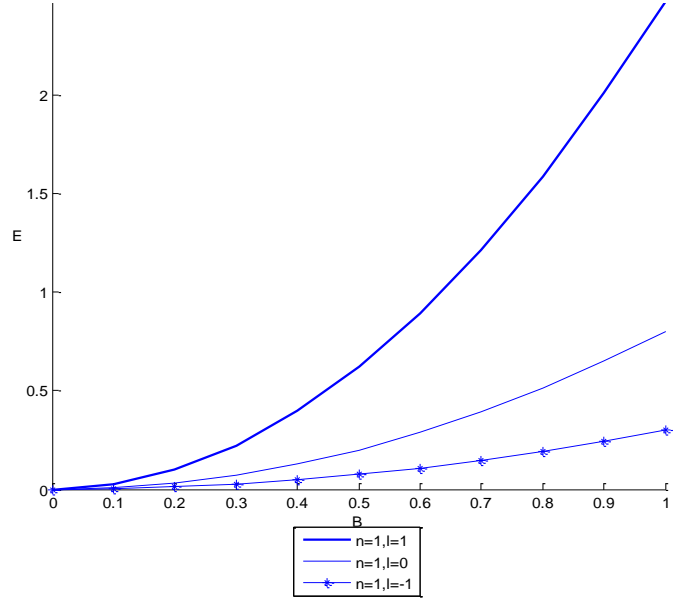

Figure 2: The energy eigenvalues versus $B(\alpha = 0.05,\ m = 1)$

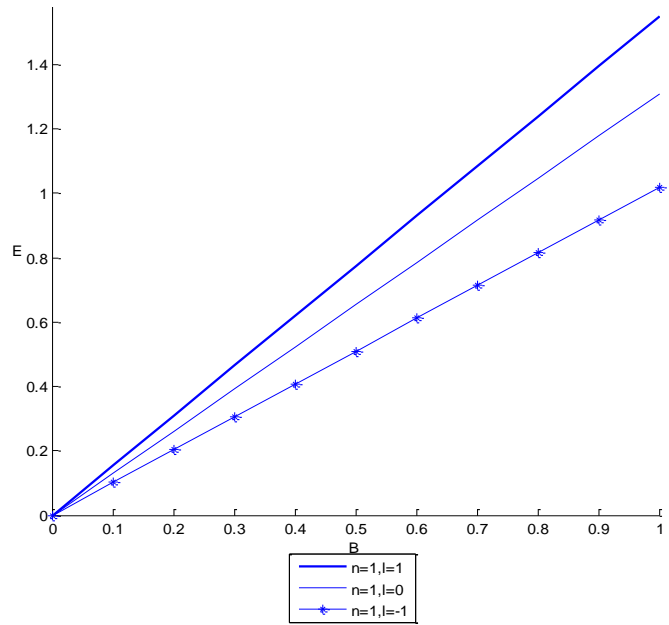

Figure 3: The energy eigenvalues versus $B$ ($m = 1$).

$$\widetilde{E}_{\tilde{n}\ell} = \pm mc^2 \left\{ \frac{\alpha\hbar^2}{m^2c^2}(4\tilde{n}^2 + 4\tilde{n}\ell + 2\ell^2)\left[m^2w^2 + \left(\frac{eB}{2c}\right)^2\right] - \frac{2\hbar mw}{m^2c^2} - \frac{\hbar\ell(eB - 2\hbar\alpha eBmw)}{m^2c^3} + 1 + \frac{2\hbar^2}{m^2c^2}(2\tilde{n} + \ell + 1)\left(m^2w^2 + \left(\frac{eB}{2c}\right)^2\right)\left(\alpha + \alpha\sqrt{1 - \widetilde{K''}}\right) \right\}^{\frac{1}{2}}, \tag{53}$$

where

$$\widetilde{K}'' = \frac{1 - 2\hbar\alpha mw - \frac{2\hbar^2\alpha^2 eBmwl}{c} - l^2\alpha^2\hbar^2\left(m^2w^2 + \left(\frac{eB}{2c}\right)^2\right)}{\alpha^2\hbar^2\left[m^2w^2 + \left(\frac{eB}{2c}\right)^2\right]}. \tag{54}$$

From the result shown in Figure 2, we see that the energy eigenvalues also increase monotonically with the magnetic field variable B, and the profile shows that the energy eigenvalues first has a slow-growth and then rapidly increases. Finally, when both the noncommutative and minimal length parameters are absent, i.e. $\alpha = \vartheta = 0$, the energy spectrum degrades into

$$\widetilde{E}_{\widetilde{n}\ell} = \pm mc^2\left[1 - \frac{\hbar\ell eB}{m^2c^3} - \frac{2\hbar w}{mc^2} + \frac{\hbar(4\widetilde{n}+2\ell+2)(w^2+(eB/2mc)^2)^{1/2}}{mc^2}\right]^{1/2}, \tag{55}$$

obviously, it is strictly consistent with [46]. In Figure 3, we have depicted the energy values versus *B*. It is also observed that the energy E increases monotonically with the magnetic field parameter *B*, and for one principal quantum number, the energy E increases with the increase of the azimuthal quantum number.

## 5. Conclusions

This paper is devoted to study of the (2 +1)-dimensional Duffin-Kemmer-Petiau oscillator for spin 0 particle under a magnetic field in the presence of a minimal length in the NC space. We first analyze the DKP oscillator in the presence of a minimal length in NC space, by employing the momentum space representation, the energy spectrum are obtained as well as the wave functions and the corresponding probability densities of the system are reported in terms of the Jacobi polynomials. Subsequently, we generalize this quantum model into the framework of a magnetic field and report the corresponding results respectively. Finally, this quantum model for special cases are discussed and the numerical results are depicted respectively. It shows that the energy eigenvalues increase monotonically with the magnetic field variable B for the minimal length parameter and the NC parameter respectively, and for one principal quantum number, the energy E increases with the increase of the azimuthal quantum number.

**Conflict of Interests**

The authors declare that there is no conflict of interests regarding the publication of this paper.

**Acknowledgments**

This work is supported by the National Natural Science Foundation of China (Grant nos. 11465006 and 11565009).